\documentclass[12pt]{article}

\begin{document}

\title{
On the Influence of Einstein--Podolsky--Rosen (EPR)
Effect on the Probability of Domain-Wall Formation
during a Cosmological Phase Transition
}

\author{
Yu.V.~Dumin\thanks{\it
IZMIRAN, Russian Academy of Sciences,
Troitsk, Moscow region, 142190 Russia,
\mbox{E-mail:~dumin@cityline.ru}
}
}

\date{}

\maketitle

\abstract{
Evading formation of the domain walls in cosmological
phase transitions is one of the key problems to be solved for
getting agreement with the observed large-scale homogeneity of
the Universe.
The previous attempts to get around this obstacle
led to imposing severe observational constraints on
the parameters of the fields involved.
Our aim is to show that yet another way
to overcome the above problem is accounting for EPR effect.
Namely, if the scalar (Higgs) field was presented by
a single quantum state at the initial instant of time,
then its reduction during a phase transition at some later instant
should be correlated even at distances
exceeding the local cosmological horizon.
By considering a simplest 1D model with $Z_2$ Higgs field,
we demonstrate that EPR effect really can substantially reduce
the probability of spontaneous creation of the domain walls.
}

\section{
Introduction
}

The problem of domain-wall formation during spontaneous breaking of
discrete symmetry was emphasized for the first time by
Zel'dovich, Kobzarev, and Okun as early as
1974--1975~\cite{Zeldovich}, i.e., just when the role of Higgs fields
in cosmology began to be recognized. This problem is associated with
the fact that the observed region of the Universe contains
a great number of domains that were not causally-connected
at the instant of phase transition, and the stable vacuum states
of such domains after the symmetry breaking will differ from
each other. As a result, a network of domain walls,
involving considerable energy density, should be formed.
But, on the other hand, the presence of such domain walls is
incompatible with the observed homogeneity of the Universe.

The previously-undertaken attempts to resolve the above problem
were based on imposing severe constraints on the parameters or
modifying the field theories involved
(see, for example,~\cite{Larsson}).

Aim of the present report is to show that, in principle,
the domain-wall problem may be resolved by a natural way
if we take into account the fact that Higgs condensate represents
a single quantum state, which should experience
Einstein--Podolsky--Rosen (EPR) correlations~\cite{Einstein}
during its reduction to the state of broken symmetry.

This point of view seems to be especially attractive due to
the recent experimental achievements, such as
(a)~the quantum-optical experiments, which confirmed a presence of
EPR correlations of the single photon states at
considerable distances ($\sim$10 km), and
(b)~the studies of Bose--Einstein condensate of ultracooled gases,
which demonstrated that all predictions of
the ``orthodox'' quantum mechanics are valid for a single quantum state
involving even a macroscopic amount of substance.

So, if we believe that EPR correlations really take place in
Higgs condensate, then it should be expected that
the probabilities of various realizations of
the Higgs-field configurations after the symmetry breaking
will be distributed by Gibbs law. As a result,
the high energy concentrated in the domain walls
(and, therefore, contradicting the astronomical observations)
turns out to be just the reason why probability of the respective
configurations is strongly suppressed.

The basic question arising here concerns the efficiency of
such suppression for a particular set of parameters of the Higgs field
under consideration. The same question can be reformulated by
an opposite way:
What parameters should the Higgs field
(and its phase transition) have for the probability of
domain-wall formation to be substantially reduced?
We shall try to give a quantitative answer to this question
in the next section.

\section{
The Model of Phase Transition\\
Allowing for EPR Correlations
}

Let us consider the simplest one-dimensional cosmological model
with metric
\begin{equation}
ds^2 = \: dt^2 - a^2(t) \, dx^2 .
\label{init_metric}
\end{equation}
By introducing the conformal time $ \eta = \int dt / a(t) $,
expression (\ref{init_metric}) can be reduced to the form
$ \, ds^2 = \, a^2(t) \: \{ d{\eta}^2  - dx^2 \}; \, $
so that the light rays
($ ds^2 = \, 0 $)
are described as
$ \, x = \, \pm \, {\eta} \, + \, {\rm const}$~\cite{Misner}.

As a result, the observed region of the Universe will contain
$N$ domains that were causally-disconnected
at the instant of phase transition:
\begin{equation}
N = \,
  ( {\eta} - {\eta}_{\scriptscriptstyle 0} ) /
    {\eta}_{\scriptscriptstyle 0} \: \approx \:
  {\eta} \, / {\eta}_{\scriptscriptstyle 0} \: ,
\end{equation}
where~$ {\eta}_{\scriptscriptstyle 0} $ is the conformal time
corresponding to the phase transition.

The Higgs field $\varphi$, possessing the symmetry group $Z_2$,
after reduction to the state of broken symmetry can be
in one of two stable vacua
($ +{\varphi}_{\scriptscriptstyle 0} \, $ or
$ \, -{\varphi}_{\scriptscriptstyle 0} $)
in each of the above-mentioned domains.
The energy of a ``wall'' between two domains with different vacua
will be denoted by $E$.
Besides, for the sake of self-consistency,
periodic boundary conditions will be imposed at
the opposite sides of the observable region;
so that the possible total number of the domain walls
is always even.

So, under assumptions formulated above,
the probability of realization of the Higgs-field configuration
involving $2k$ domain walls equals
\begin{equation}
P^{2k}_{\scriptscriptstyle N} \, = \:
  2 \, A_{\scriptscriptstyle N} \,
  \frac{N!}{(2k)! \; (N \! - \! 2k)!} \: e^{-2kE/T} ,
\end{equation}
where $T$ is the temperature of phase transition, and
$A_{\scriptscriptstyle N}$ is the normalization factor,
defined as
\begin{equation}
A^{\scriptscriptstyle -1}_{\scriptscriptstyle N} \, = \:
  \sum^{\scriptscriptstyle [N/2]}_{k=0} \:
  \frac{2 \: N!}{(2k)! \; (N \! - \! 2k)!} \: e^{-2kE/T} ,
\label{norm_factor}
\end{equation}
where square brackets in the upper limit of the sum denote
the integer part of a number.

Normalization factor~(\ref{norm_factor}) can be easily calculated
in two limiting cases:
(a)~when neighboring terms of the sum differ from each other
only slightly, or
(b)~when main contribution to the sum is done by a few first terms.

As regards the first case,
summation can be approximately extended to the terms with any number
of the domain walls (both odd and even) and finally gives
\begin{equation}
A^{\scriptscriptstyle -1}_{\scriptscriptstyle N} \, \approx \:
  {\left( 1 + e^{-E/T} \right)}^{\scriptscriptstyle N} ;
\end{equation}
so that
\begin{equation}
P^{2k}_{\scriptscriptstyle N} \, \approx \:
  \frac{2 \: N!}{(2k)! \; (N \! - \! 2k)!} \;
  \frac{e^{-2kE/T}}%
  {{\left( 1 + e^{-E/T} \right)}^{\scriptscriptstyle N}} \; .
\end{equation}

In particular, probability of the Higgs-field configuration without
any domain walls (which is just the case actually observed) equals
\begin{equation}
P^{0}_{\scriptscriptstyle N} \, \approx \:
  \frac{2}{{\left( 1 + e^{-E/T} \right)}^{\scriptscriptstyle N}} \; .
\end{equation}

It is interesting to find restriction on the domain-wall energy $E$
and phase-transition temperature $T$ for
the probability of absence of the domain walls
to be greater than or equal to some specified number~$p$
($ 0 < p < 1 $, for example, $ \, p = 1/2 $):
\begin{equation}
E/T \, \ge \,
  \ln \frac{1}{(2/p)^{\scriptscriptstyle 1/N} - 1} \, \approx \,
  \ln N - \ln \ln \, (2/p) \: \approx \:
  \ln N \, .
\label{E-T_high_temp}
\end{equation}

As regards the opposite case,
when normalization factor~(\ref{norm_factor}) is determined by
the terms with small~$k$, it can be estimated by taking into account
only the first two terms:
\begin{equation}
A^{\scriptscriptstyle -1}_{\scriptscriptstyle N} \, \approx \:
  2 \: \left\{ 1 + \frac{1}{2} \, N^{2} \, e^{-2E/T} \right\} ;
\end{equation}
and the respective probability of the Higgs-field configuration
involving $2k$ domain walls is
\begin{equation}
P^{2k}_{\scriptscriptstyle N} \, \approx \:
  \left\{ 1 - \frac{1}{2} \, N^{2} \, e^{-2E/T} \right\}
  \frac{N!}{(2k)! \; (N \! - \! 2k)!} \; e^{-2kE/T} .
\end{equation}

Then, the probability that domain walls are absent at all is
\begin{equation}
P^{0}_{\scriptscriptstyle N} \, \approx \:
  1 - \frac{1}{2} \, N^{2} \, e^{-2E/T} ,
\end{equation}
and it will be greater than or equal to the specified number~$p$ if
\begin{equation}
E/T \, \ge \,
  \ln N - \frac{1}{2}  \, \ln \, ( 2 \, (1 \!\! - \! p) ) \: \approx \:
  \ln N \, ,
\label{E-T_low_temp}
\end{equation}
i.e., to a first approximation, it is again determined by
the logarithmic function of $N$.

\section{
Conclusion
}

There is no doubt that a considerably more sophisticated analysis
must be carried out to draw unambiguous conclusion on the role of
EPR correlations in the phase transitions of Higgs fields.
Nevertheless, our calculations, based on the simplest model,
showed that such possibility is quite promising.
Since, as follows from~(\ref{E-T_high_temp})
and~(\ref{E-T_low_temp}),
the ratio $ E/T $ differs from unity only by $ \ln N $,
then the required temperature of phase transition is
of the same order of magnitude as the domain-wall energy
even at a large number of the causally-disconnected regions~$N$.
Therefore, the required parameters of the phase transition
could be satisfied in some particular kind of the field theory.

\end{document}